# Wavelength-resolved polarization characterization of supercontinuum vector beams

SHUNPEI NARA,[1] HAYATO KOBAYASHI,[1] KOUSUKE KUSANO,[1] YUTA TAKAI,[1] HIROYUKI KOWA,[2] TOSHITAKA WAKAYAMA,[3] AND TAKESHI HIGASHIGUCHI[1,*]

[1]*Department of Electrical and Electronic Engineering, Faculty of Engineering, Utsunomiya University, 7-1-2 Yoto, Utsunomiya, Tochigi 321-8585, Japan*
[2]*EMT Co., Ltd. 5-10-4-101, Ozakudai, Hamura, Tokyo 205-0001, Japan*
[3]*Faculty of Health and Medical Care, Saitama Medical University, 1397-1 Yamane, Hidaka, Saitama 350-1241, Japan*
*hgsgc_tks@me.com

**Abstract:** The polarization state of a supercontinuum (SC) vector beam provides a reference for quantitative polarimetry. We characterize an azimuthally polarized SC vector beam generated with an S-waveplate and azimuthal polarizers by wavelength-resolved Stokes imaging. Representative maps show an approximately tangential azimuthal angle and a small ellipticity parameter over much of the beam, together with the degree of polarization. The ± 90° azimuthal angle boundary shifts by about 21° between 450 and 766 nm. A linear fit gives 21° ± 7°, where the uncertainty is derived from the slope confidence interval and does not include uncharacterized systematic errors. Jones-matrix calculations examine candidate imperfections under specified assumptions. Attributing the measured boundary shift to this mechanism requires an intensity extinction ratio of about 7 or less at 766 nm if the offset at 450 nm is negligible and the shift corresponds to a uniform azimuthal angle change. The origin remains unresolved. The results motivate wavelength-dependent characterization and calibration of this probe for quantitative polarimetry.

## 1. Introduction

Supercontinuum (SC) light is generated by nonlinear processes in photonic crystal fibers and other waveguides, and its spectrum extends from the visible to the near infrared. It is used in spectroscopy, optical frequency metrology, imaging, and nonlinear optics. Soliton fission, dispersive-wave generation, Raman shift, and coherence in SC generation, together with the guiding properties of photonic crystal fibers, have been studied extensively [1-6]. Optical vortices [7,8] and vector beams, such as radially and azimuthally polarized beams [9], are structured light in which the phase or the polarization is controlled across the beam cross section. The spatially varying polarization of vector beams has been applied to the control of tightly focused fields [10,11], optical trapping, super-resolution microscopy, and laser processing [9,12]. An SC vector beam combines a broad spectrum with a spatial polarization structure. Because the birefringence and optical anisotropy of a sample depend on wavelength, such a beam can record the polarization response at many wavelengths in a single measurement, which single-wavelength and wavelength-scanning methods cannot do.

Space-variant waveplates based on the geometric phase [13,14] have been developed [15], and *q*-plates have been used to control the polarization and orbital angular momentum of broadband light [16,17]. We previously proposed a method that controls the polarization and the geometric phase of SC vector beams independently, and generated azimuthally and radially polarized beams with it [18]. Using the spatial polarization structure of these beams, we then demonstrated single-shot multispectral birefringence mapping and the detection of birefringence singularities [19,20]. Multiwavelength birefringence evaluation is also relevant to stress management in optical elements and to functional elements, for example, high-sensitivity mapping of stress birefringence in fused silica substrates by cavity ring-down [21] and waveplates written inside silicon through stress-induced birefringence [22].

The spatial, spectral, and polarization degrees of freedom of broadband structured light can be correlated, and such correlations can be engineered [23]. Wavelength-dependent polarization distributions have also been investigated in cylindrically polarized pulses [24] and in vector supercontinuum generated by filamentation [25]. Spectroscopic polarization measurements with a time-encoded SC vector beam have been demonstrated [26]. These studies establish the importance of spectrally resolved polarization characterization and measurement, although their generation or measurement schemes differ. For the specific S-waveplate and azimuthal-polarizer arrangement used here, the wavelength dependence of the spatial polarization state and the possible influence of component imperfections require further characterization. We combine spatially resolved Stokes measurements at visible wavelengths with a quantitative boundary-angle analysis and a Jones-matrix examination of candidate imperfections. This combination characterizes the wavelength-dependent behavior of the present implementation. In birefringence measurements, the probe polarization provides a reference for retrieving sample properties [27]. A wavelength-dependent change of that reference can bias the inferred principal-axis orientation if it is not calibrated.

Here we characterize the wavelength-resolved polarization state of an azimuthally polarized SC vector beam generated with an S-waveplate and azimuthal polarizers. We examine the spatial distributions of intensity, degree of polarization, ellipticity, and azimuthal angle at representative wavelengths. We then quantify the wavelength dependence of the $\pm$ 90° boundary in the azimuthal angle maps. Finally, we analyze the Stokes measurement and use Jones-matrix calculations to examine whether selected nonidealities can produce an azimuthal angle change comparable to the observed boundary shift. The calculations provide possible mechanisms under stated assumptions, rather than an experimental identification of the responsible element. The measured boundary shift illustrates why the polarization reference should be checked at each wavelength before this beam is used for quantitative multiwavelength polarimetry. The magnitude and spatial uniformity of the associated azimuthal angle change, and its effect on a particular sample reconstruction, require further validation. We characterize the observed change and examine its possible origin without assuming that the cause has been established.

## 2. Experimental methods

### *2.1 Experimental setup*

Figure 1(a) shows the setup for generating the SC vector beam. It consists of the SC source, an optical system for optical vortex generation, and an optical system for azimuthal polarization generation. Light from the SC source is limited to the visible range by an 800 nm short-pass filter, and the beam diameter is enlarged with a beam expander. A polarizing beam splitter (PBS) converts the light to linear polarization, and a QWP converts it to circular polarization. The S-waveplate then generates an optical vortex. It is an optical element with a fine periodic structure, designed so that the slow-axis orientation changes continuously with axial symmetry [28-30]. A second QWP converts the circularly polarized vortex to linear polarization, and a second PBS transmits only the linear component that carries the vortex phase and rejects the residual component that the S-waveplate leaves unconverted. A third QWP restores circular polarization, and the azimuthal polarizer converts the vortex into azimuthal polarization. Two azimuthal polarizers are stacked to improve the extinction ratio.

Figure 1(b) shows the spectrum of the SC source after the short-pass filter. The intensity rises near 450 nm and stays high up to 800 nm, beyond which the filter suppresses it. The 450

– 800 nm band therefore has sufficient intensity for polarization analysis. The azimuthal angle was evaluated at 400 – 766 nm, but at 400 nm the intensity was too low for a clear evaluation.

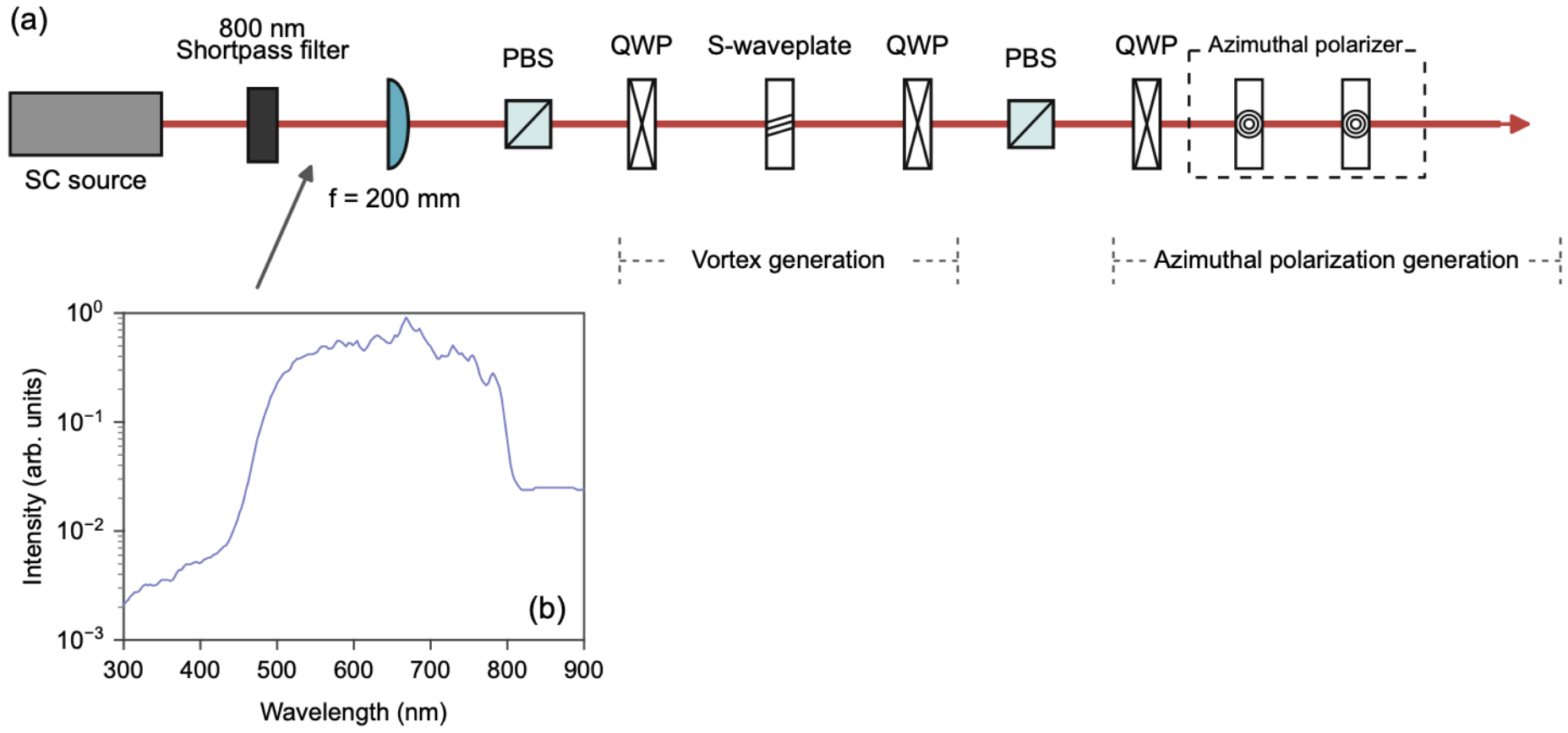

Fig. 1. (a) Optical setup for generating the azimuthally polarized SC vector beam. An 800 nm short-pass filter limits the spectrum, and a beam expander and a lens with a focal length of 200 mm follow. The optical vortex generation and azimuthal polarization generation stages are indicated by dashed brackets. The conversion system is described in Section 2.1 and Ref. [18]. (b) Spectrum of the SC source after the short-pass filter on a logarithmic intensity axis in arbitrary units. The signal rises near 450 nm and remains high up to 800 nm. The 400 nm signal was too low for reliable polarization analysis.

The polarization conversion in Fig. 1(a) follows the scheme in Ref. [18], in which the vortex phases of the left- and right-handed components cancel. The QWPs are assumed to be achromatic and the polarizers ideal, $\theta$ is the azimuthal coordinate in the beam cross section, and all orientations are measured from the transmission axis of the first PBS (0). The first QWP (fast-axis orientation at $\pi/4$) converts the linearly polarized light into circular polarization. The S-waveplate is a half-wave retarder whose slow-axis orientation is $\theta/2$ and whose retardance is $\Delta = \pi + \varepsilon$. Assuming an approximately wavelength-independent optical path difference, $\varepsilon \approx \pi(\lambda_0/\lambda - 1)$, where $\lambda_0$ is the wavelength at which $\Delta = \pi$ (taken as 532 nm for the numerical examples). The S-waveplate converts a fraction $\cos^2(\varepsilon/2)$ of the power into the circular polarization of opposite handedness that carries the geometric phase $e^{-i\theta}$, that is, an optical vortex, whereas the remaining fraction $\sin^2(\varepsilon/2)$ stays in the incident circular state without the vortex phase. The coherent sum of the two components gives $S_3/S_0 = -\cos\varepsilon$ and a degree of linear polarization $P_L = |\sin\varepsilon|$ immediately after the S-waveplate; the azimuthal angle of this weak linear component follows $\theta/2$ plus a constant. The second QWP (fast-axis orientation at $-\pi/4$) maps the two circular components onto orthogonal linear polarizations, and the second PBS, whose transmission axis ($\pi/2$) is orthogonal to that of the first, transmits only the vortex component and rejects the residual one. The third QWP ($\pi/4$) restores circular polarization, and the azimuthal polarizer, whose local transmission axis is $\theta + \pi/2$, projects this light onto the local azimuthal direction. The projection imparts the opposite geometric phase $e^{+i\theta}$, which cancels the vortex phase imparted by the S-waveplate. The output Jones vector is therefore proportional to $\cos(\varepsilon/2)(-\sin\theta, \cos\theta)^{\mathrm{T}}$ with a spatially uniform phase, and the output Stokes vector, normalized to the intensity after the first PBS, is $(1/2)\cos^2(\varepsilon/2)(1, -\cos 2\theta, -\sin 2\theta, 0)^{\mathrm{T}}$. For ideal elements, the azimuthal angle of the polarization is thus $\varphi = \theta + \pi/2$ and the degree of polarization is $P = 1$ at every wavelength. The retardance error $\varepsilon$ neither rotates the azimuthal angle nor distorts the phase. It only lowers the output intensity through the factor $\cos^2(\varepsilon/2)$, which stays at or above 0.79 over 450 – 766 nm for $\lambda_0 = 532$ nm. In a geometric-phase retarder used without such filtering, a retardance error also distorts the ellipticity and the intensity distribution [31]. Here the second PBS removes the unconverted component. A wavelength-dependent rotation of $\varphi$ at the output would indicate a departure from this idealized generation model, while measurement errors can also affect the observed azimuthal angle.

### 2.2 Evaluation method

To confirm that the vector beam has the properties of azimuthal polarization, we evaluated the spatial distributions of the degree of polarization, the ellipticity, and the azimuthal angle of the polarization in Fig. 2. All three were calculated from the Stokes parameters ($S_0$, $S_1$, $S_2$, $S_3$). The degree of polarization $P$ is

$$P = \frac{\sqrt{S_1^2+S_2^2+S_3^2}}{S_0}, \tag{1}$$

and is equal to 1 for fully polarized light. We define the ellipticity parameter $\chi$ as

$$\chi = \frac{S_3}{S_0+\sqrt{S_1^2+S_2^2}}. \tag{2}$$

For fully polarized light, $\chi$ is the signed minor-to-major axis ratio of the polarization ellipse. For partially polarized light, Eq. (2) is a Stokes-derived parameter and need not equal the ellipticity of the polarized component. The azimuthal angle $\varphi$ is calculated with quadrant correction as one half of $\tan^{-1}(S_2/S_1)$, modulo 180°, corresponding to

$$\varphi = \frac{1}{2}\tan^{-1}\left(\frac{S_2}{S_1}\right). \tag{3}$$

The Stokes parameters were obtained with a rotating QWP and a fixed linear polarizer (LP), a standard configuration whose design and systematic errors have been analyzed in detail [32]. A camera recorded images at 36 QWP orientations in 10° steps, and the Stokes parameters were calculated from them. The detected intensity, as a function of the fast-axis orientation $\beta$ of the QWP, is the Fourier series

$$I(\beta) = a_1 + a_2\sin 2\beta + a_3\cos 4\beta + a_4\sin 4\beta\ , \tag{4}$$

and the Stokes parameters follow from the Fourier coefficients as

$$S_0 = 2(a_1 - a_3),\ S_1 = 4a_3,\ S_2 = 4a_4,\ S_3 = -2a_2\ . \tag{5}$$

Measurements were not repeated under identical conditions. The polarization dependence of the fixed LP and the camera was not evaluated. According to the manufacturer's specifications, the retardance of the achromatic quarter-wave plate deviates from 90° by no more than 3° over the wavelength range of 450 – 700 nm.

## 3. Results and discussion

### 3.1 Polarization parameters of the SC vector beam

Figure 2 summarizes the polarization parameters measured over the visible spectrum without wavelength selection. The degree of polarization in Fig. 2(a) averages about 0.64 over the beam profile, and the mean ellipticity in Fig. 2(b) is about 0.078. The azimuthal angle in Fig. 2(c) changes continuously around the beam center and is approximately perpendicular to the radius. The jump between + 90° and – 90° reflects the angular convention. The polarization-ellipse display in Fig. 2(d) shows a doughnut-shaped intensity distribution with approximately tangential, nearly linear polarization. These broadband quantities integrate spectral contributions and should not be interpreted as the polarization parameters at an individual wavelength.

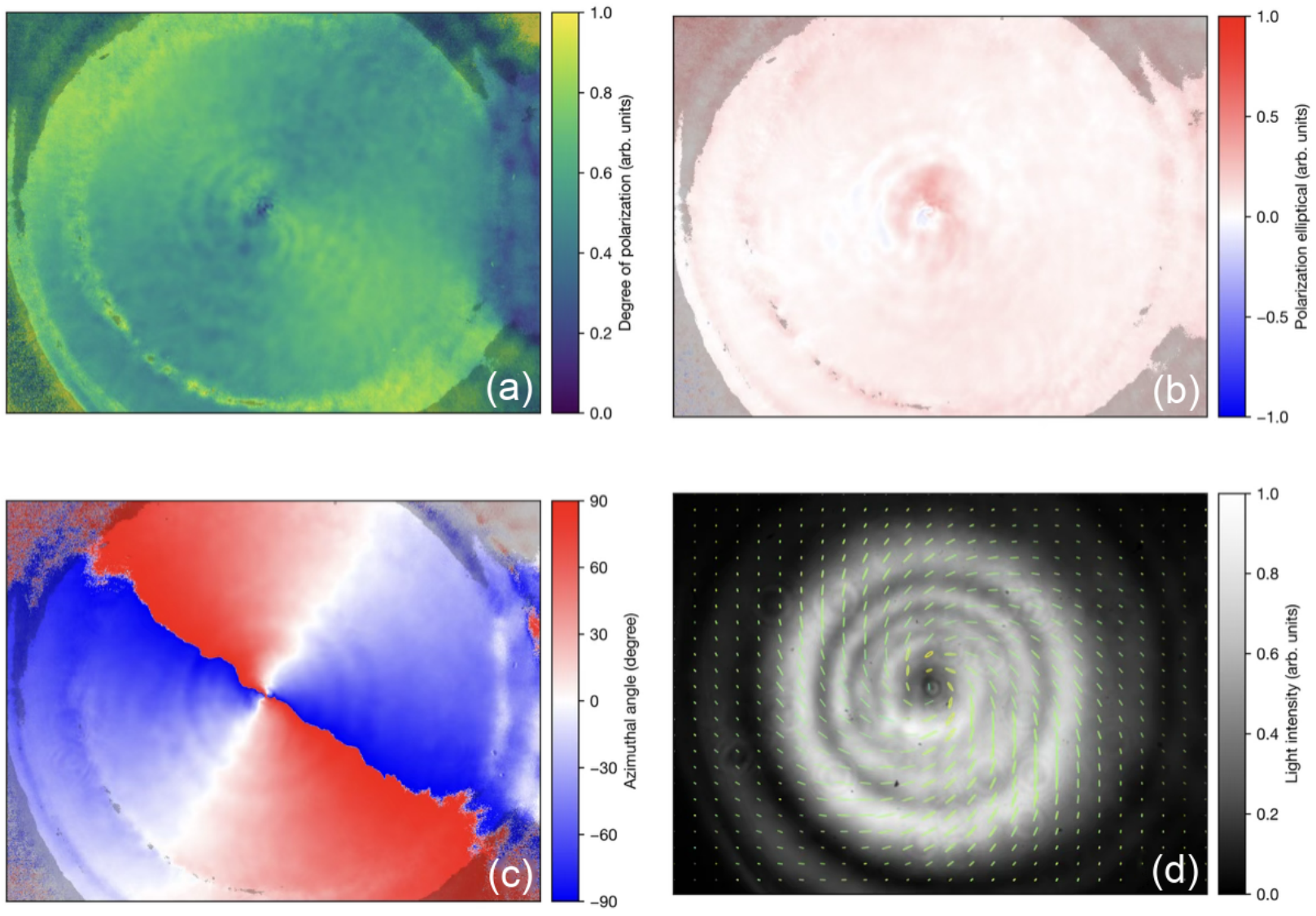

Fig. 2. Spatial distributions of the polarization parameters of the SC vector beam, measured over the full visible range without wavelength selection. (a) Degree of polarization $P$ (color scale, 0 – 1). Its average over the beam profile is about 0.64. (b) Ellipticity $\chi$ (color scale, – 1 to 1; red, positive; blue, negative). Its average over the beam profile is about 0.078. (c) Azimuthal angle $\varphi$ (color scale, – 90° to + 90°). The sharp boundary between red and blue is not a physical discontinuity; it appears because $\varphi$ is defined through arctan and wraps from + 90° to – 90°. (d) Polarization ellipses superimposed on the intensity image. The direction of each line shows the azimuthal angle of the polarization, its shape shows the ellipticity (line, linear; ellipse, elliptical; circle, circular), and its color shows the polarization state (red, left-handed circular; blue, right-handed circular; green, linear).

Figure 3 shows the wavelength-resolved intensity and polarization distributions at 450, 532, 632, and 766 nm. The same 10 nm bandwidth was used at each wavelength. The spatial azimuthal angle pattern is approximately tangential, and the ellipticity is generally close to zero in the illuminated region. The degree-of-polarization maps are less uniform at 450 nm than at the higher-intensity representative wavelengths. These observations are qualitative, particularly near the beam center and in low-intensity regions. The 400 nm measurement was too weak for a clear evaluation. The wavelength-resolved distributions differ from the broadband averages in Fig. 2, which combine light from the full passband.

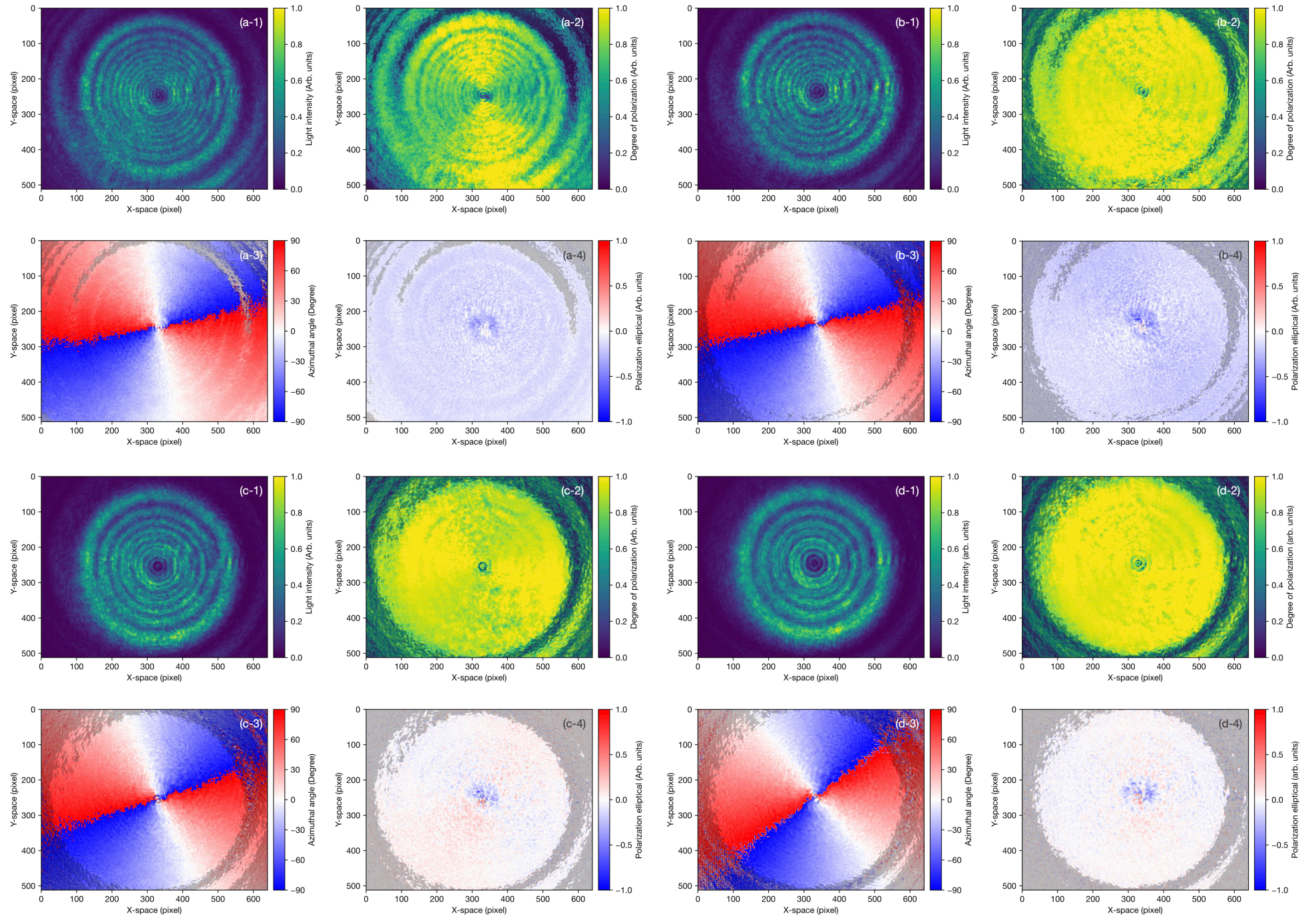

Fig. 3. Spatial distributions of intensity $S_0$, degree of polarization P, azimuthal angle φ, and ellipticity χ at (a) 450 nm, (b) 532 nm, (c) 632 nm, and (d) 766 nm. Each set of four panels shows (a-1) – (d-1) intensity at upper left, (a-2) – (d-2) degree of polarization at upper right, (a-3) – (d-3) azimuthal angle at lower left, and (a-4) – (d-4) ellipticity at lower right. A 10 nm bandpass filter was used for each measurement. The intensity is in arbitrary units, and the axes are camera pixel coordinates. Low-intensity regions, particularly near the beam center and the beam edge, should be interpreted with caution.

### *3.2 Wavelength dependence of the azimuthal angle distribution*

To examine the wavelength dependence of the azimuthal angle in more detail, we inserted a bandpass filter (FKSPVIS, Thorlabs; bandwidth 10 nm) into the vector-beam path and used the same Stokes imaging procedure as for Fig. 2. Figures 4(a) – 4(d) show representative azimuthal angle maps at 589, 632, 660, and 700 nm. The narrower passband reduces the signal, and the maps are noisier than the broadband image. The ± 90° boundary moves counterclockwise in the displayed coordinates as wavelength increases. We expanded the beam profiles in polar coordinates about the beam center and plotted the azimuthal angle against polar angle in Fig. 4(e). The discontinuities of this plot provide a way to locate the boundary. At 400 nm the signal was insufficient for a clear evaluation. At the end wavelengths some discontinuities split into several lines, which makes their positions uncertain. The boundary shift is quantified in Section 3.3. It does not by itself establish a spatially uniform rotation of the full azimuthal angle distribution.

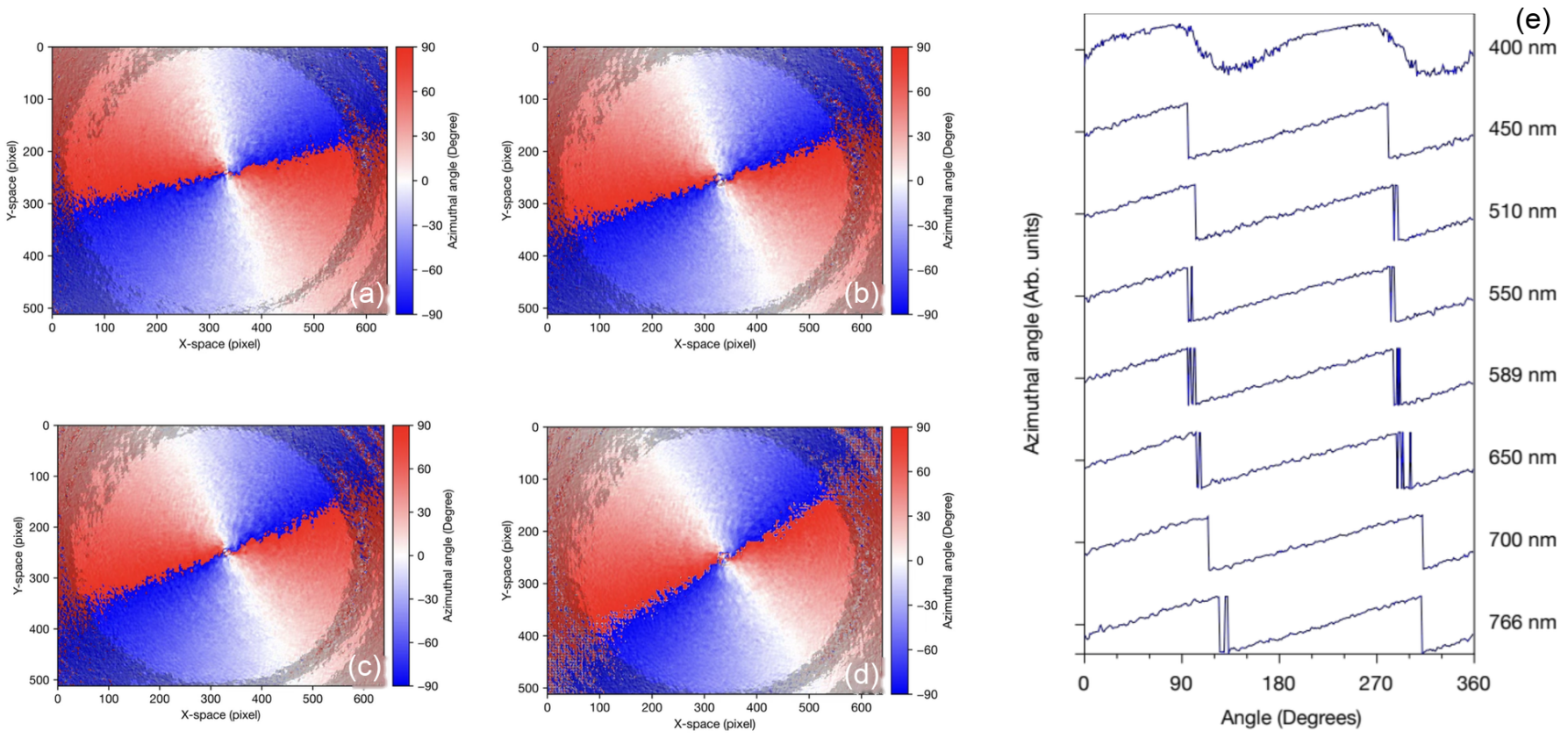

Fig. 4. Wavelength dependence of the azimuthal angle distribution. (a) – (d) Azimuthal angle maps measured through 10 nm bandpass filters centered at 589, 632, 660, and 700 nm, respectively. The color scale spans – 90° to + 90°, and the axes are camera pixel coordinates. (e) Azimuthal angle plotted against polar angle for the indicated wavelengths after expanding the beam profiles about the beam center. Curves are offset vertically for clarity. The angular discontinuities locate the ± 90° boundary and shift with wavelength. At 400 nm the low signal prevents a clear boundary evaluation. Split or broadened discontinuities at the ends of the wavelength range make their locations uncertain.

### *3.3 Wavelength dependence of the boundary angle*

Figure 5(a) illustrates the boundary-angle measurement at 700 nm. At each wavelength, the angle between the horizontal line through the beam center and the ± 90° boundary was estimated using the discontinuities in Fig. 4(e). Figure 5(b) shows an increase from about 11° at 450 nm to about 32° at 766 nm. The reported linear regression has a slope of (0.066° ± 0.021°)/nm, a correlation coefficient of 0.89, and a residual standard deviation of 3.2°. The slope gives a fitted boundary-angle change of 21 ± 7° over 450 – 766 nm, with the quoted uncertainty derived from the reported 95% slope interval. The plotted ± 3.2° error bars represent the fit residual standard deviation, not repeated-measurement uncertainties or a full systematic-error budget. The trend is not exactly linear. The boundary angle remains about 9° – 16° below 590 nm and increases more steeply above 630 nm.

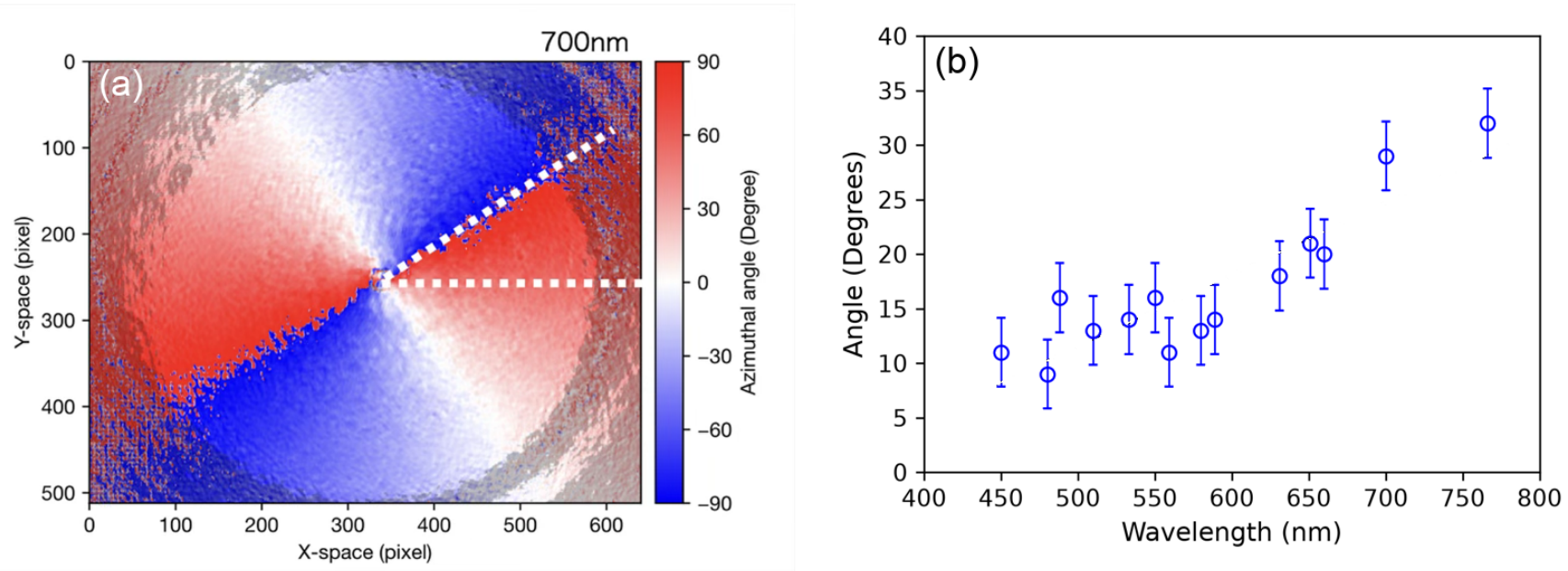

Fig. 5. (a) Measurement of the angle between the horizontal line through the beam center and the ±90° azimuthal angle boundary, illustrated for 700 nm. The color scale spans – 90° to + 90°. (b) Boundary angle versus filter center wavelength. The open circles show the measured values and the error bars show ± 3.2°, the residual standard deviation of the linear fit. The reported slope is (0.066° ± 0.021°)/nm at the 95% confidence level, corresponding to a fitted change of 21° ± 7° over 450 – 766 nm. These bars do not represent a repeatability measurement or include all systematic errors.

### *3.4 Possible origin of the wavelength dependence*

The ± 90° boundary changes by about 21° between 450 and 766 nm. We examine whether modeled imperfections can produce an azimuthal angle change of comparable magnitude. A fixed alignment offset may contribute to the absolute boundary angle, so we focus on its change with wavelength. The boundary angle and the spatial mean of the local azimuthal angle deviation are different observables. Comparing them assumes that the deviation is approximately uniform across the analyzed region, an assumption that has not been established by the boundary measurement alone. For a uniform offset $\varphi = \theta + \pi/2 + \alpha$, a boundary shift $\Delta\theta$ is opposite to the change in $\alpha$ when both angles use the same counterclockwise convention. The reported direction therefore describes the boundary, not necessarily the local polarization rotation. For ideal elements, the output azimuthal angle is $\theta + \pi/2$ and the S-waveplate retardance error changes only the intensity, as described in Section 2.1.

We consider the measurement system. The azimuthal angle is obtained from one half of the two-argument arctangent of $S_2$ and $S_1$, so polarization-independent attenuation does not change it. In the ideal rotating-QWP and fixed-analyzer model, a QWP with retardance $\delta$ multiplies the fourth-harmonic coefficients associated with $S_1$ and $S_2$ by the same factor $(1 - \cos\delta)/4$. Their ratio, and hence the retrieved azimuthal angle, is therefore unaffected by the retardance error alone [33]. This result assumes the ideal analyzer model, correct angular reference, and no additional polarization-dependent response. Misalignment together with wavelength-dependent retardance, nonideal optics, detector response, and other systematic errors have not been independently characterized. The analysis therefore does not rule out a wavelength-dependent measurement artifact. We next use Jones matrices to test selected errors upstream of the azimuthal polarizer. For an ideal second PBS, its transmitted field has one linear polarization, so imperfections before it change the transmitted amplitude rather than the state incident on the third QWP. A finite PBS extinction ratio relaxes this conclusion. The calculations below examine error values and combinations, not all possible upstream nonidealities. Achromatic QWPs can retain wavelength-dependent retardance errors [34]. The third QWP may therefore produce elliptically rather than perfectly circularly polarized light at the azimuthal polarizer. We write its retardance as $\pi/2 + d(\lambda)$ and take its fast-axis orientation to be $\pi/4$ in the model. The corresponding input Stokes vector, normalized to the input intensity, is $(1, \sin d, 0, -\cos d)$. The residual linear component has magnitude $|\sin d|$.

The azimuthal polarizer may have a finite extinction ratio. We therefore treat the output as azimuthal polarization with a superposed radially polarized component. In the calculation, the polarizer is a partial polarizer with an amplitude transmittance of 1 along the local azimuthal direction $\theta + \pi/2$ and $r$ $(0 \le r < 1)$ along the radial direction $\theta$. In this model there is no retardance between the two components. The intensity extinction ratio is $1/r^2$. For stacked polarizers with aligned axes, $r$ is the product of the individual amplitude ratios. The output Stokes vector, normalized to the intensity after the first PBS as in Section 2.1, is

$$S_0 = \frac{1}{2}\cos^2(\varepsilon/2)[(1 + r^2) - (1 - r^2)\sin d \cos 2\theta], \tag{6a}$$

$$S_1 = -\frac{1}{2}\cos^2\left(\frac{\varepsilon}{2}\right)\left[(1 - r^2)\cos 2\theta - \frac{(1+r)^2}{2}\sin d - \frac{(1-r)^2}{2}\sin d \cos 4\theta\right], \tag{6b}$$

$$S_2 = -\frac{1}{2}\cos^2\left(\frac{\varepsilon}{2}\right)\left[(1 - r^2)\sin 2\theta - \frac{(1-r)^2}{2}\sin d \sin 4\theta\right], \tag{6c}$$

$$S_3 = -r\cos^2(\varepsilon/2)\cos d. \tag{6d}$$

For $r = 0$ and $d = 0$, Eq. (6) reduces to the ideal result in Section 2.1. First, the retardance error $\varepsilon$ of the S-waveplate enters only through the common factor $\cos^2(\varepsilon/2)$. It lowers the intensity but does not change the normalized Stokes parameters. Second, the leaked radial component gives a nonzero $S_3 = -r\cos^2(\varepsilon/2)\cos d$. For small $r$ and $d$, neglecting the angular correction of order $rd$, Eq. (2) gives the approximation

$$\chi \approx -r\cos d, \tag{7}$$

In this approximation, $|\chi| \approx r|\cos d| \approx r$. At finite $d$, the first-order expression in $r$ also depends on $\theta$, so Eq. (7) should not be used without the small-$d$ assumption. The broadband mean ellipticity in Fig. 2 is not an independent measurement of $r$ at any one wavelength, and no extinction-ratio bound is inferred from it. The $S_1$ and $S_2$ terms change with the combination of $d$ and $r$. For $r = 0$, $d$ modulates the intensity but does not change the azimuthal angle. For $d = 0$, the azimuthal angle remains $\theta + \pi/2$ even though $S_3$ is nonzero. For small $d$, the deviation from $\theta + \pi/2$ is

$$\Delta\varphi \approx \frac{r}{1-r^2} d \sin 2\theta \tag{8}$$

In this model, wavelength-dependent $d$ can change the ellipticity, intensity modulation, and azimuthal angle distortion. Equation (8) vanishes at $\theta = 0$ and $\pi/2$ and has zero angular mean under uniform angular weighting. Thus, the modeled QWP error and polarizer leakage produce an angular distortion rather than a uniform rotation. This does not exclude other imperfections or establish that the measured distribution has the same form.

The numerical Jones-matrix calculations in Table 1 test the optical sequence P(0°) → QWP(π/4) → S-waveplate → QWP(−π/4) → P(π/2) → QWP(π/4) → azimuthal polarizer. The S-waveplate slow-axis orientation is $\theta/2$ and its retardance is $\pi + \varepsilon$, with $\varepsilon = \pi(532/\lambda - 1)$. The polarizer transmission-axis orientation is $\theta + \pi/2$. At 450, 500, 532, 589, 632, 700, and 766 nm, the calculation evaluates the angular mean azimuthal angle deviation and its standard deviation. For the parameter cases listed in Table 1, the mean rotation from the combined errors without polarizer retardance is at most 0.26°, with a standard deviation of about 2°. These model quantities are not identical to the measured boundary angle, and this comparison does not establish an exhaustive exclusion of upstream errors.

Table 1. Mean rotation and standard deviation of the output azimuth relative to $\theta + \pi/2$, calculated with Jones matrices for the errors listed. Each row gives the largest absolute mean rotation and the range of the standard deviation (taken over $\theta$) over the seven wavelengths from 450 to 766 nm. The error of the PBSs is a leaked amplitude of the rejected polarization, and the polarization-dependent loss is an amplitude factor on the slow axis of the S-waveplate. In the last row, the sign of the mean rotation is for the handedness of Eq. (6); the opposite handedness gives +11.3°.

| Error source | Parameter | Mean rotation (°) | Std. dev. (°) |
|---|---|---|---|
| S-waveplate, polarization-dependent loss | amplitude 0.8 | 0.00 | 0.00 |
| QWPs, retardance error (all three) | +10° | 0.00 | 0.00 |
| QWPs, fast-axis error (all three) | 5° | 0.00 | 0.00 |
| PBSs, finite extinction ratio (both) | amplitude ratio 0.1 | 0.00 | 0.00 |
| Azimuthal polarizer, leakage | $r = 0.1$ | 0.00 | 0.00 |
| Polarizer leakage and third-QWP retardance error | $r = 0.2$, $d = 10°$ | 0.00 | 1.5 |
| Polarizer leakage and wavelength-dependent third-QWP error | $r = 0.2$, $d = (\pi/2)(532/\lambda - 1)$ | 0.00 | 0.0 – 4.3 |
| All of the above combined | $r = 0.2$ | 0.26 | 1.9 – 2.1 |
| Polarizer retardance $\Gamma = \pi/2$ | $r = 0.2$ | – 11.3 | 0.00 |

The partial-polarizer model above assumes that the transmitted and leaked components are in phase. A space-variant polarizer may also introduce a retardance $\Gamma$ between the local radial and azimuthal components. For circular input, a phase shift between these components can convert part of the local circular Stokes component into a linear component. The output azimuthal angle then has the same offset at each θ in this idealized model. With the sign convention of Eq. (6),

$$\tan 2\Delta\varphi_0 = -\frac{2r\sin\Gamma}{1-r^2}, \tag{9}$$

where the sign depends on the incident handedness and the retardance convention. The output ellipticity is approximately $|\chi| \approx r|\cos\Gamma|$ for small $r$. A nonzero $S_3$ alone does not rotate the azimuthal angle in this model. A rotation occurs when leakage and retardance are both present, with a contribution proportional to $r \sin\Gamma$. At $\Gamma = \pi/2$, the ellipticity vanishes and $|\Delta\varphi_0| = \tan^{-1} r$, the largest rotation for a given $r$. If the offset at 450 nm is negligible and the whole measured boundary change is attributed to a uniform polarizer-induced azimuthal angle offset, a change of 21° requires $r \geq \tan 21° \approx 0.38$ at 766 nm, corresponding to an intensity extinction ratio of about 7 or less. This is a conditional model estimate, not a measured extinction ratio. The data do not establish whether these conditions hold. Reversing the incident circular handedness could help test the modeled mechanism, provided that the other elements and measurement system are separately characterized.

The calculations narrow the possibilities within the models and parameter values tested, but they do not locate the experimental cause. Possible contributions include the azimuthal polarizer, downstream optics, finite extinction or other imperfections in upstream elements, and the measurement system. The polarizer extinction ratio and retardance, the downstream response, and wavelength-dependent systematic errors of the polarimeter were not independently measured. Characterization of known polarization states and of individual elements would help distinguish these possibilities. Until then, wavelength-resolved calibration of the probe is important for quantitative applications.

## 4. Conclusion

We characterized an azimuthally polarized SC vector beam by wavelength-resolved Stokes imaging. At representative wavelengths from 450 to 766 nm, the maps show an approximately tangential azimuthal angle and a small ellipticity parameter over much of the beam, together with spatial variations in the degree of polarization. The ± 90° azimuthal angle boundary shifted by about 21° between 450 and 766 nm. The fitted change was 21° ± 7°, with the uncertainty derived from the slope confidence interval rather than a complete uncertainty budget. The boundary shift alone does not establish a spatially uniform rotation. Under the assumed element models, selected upstream errors and azimuthal polarizer leakage without retardance did not produce a comparable uniform mean rotation. Leakage combined with retardance can produce one, but an extinction ratio of about 7 or less at 766 nm follows only if the 450 nm offset is negligible and the boundary shift is attributed to a uniform offset from this mechanism. The physical origin and the contribution of systematic measurement errors remain unresolved.

The results show that the polarization reference of this SC vector beam should not be assumed wavelength invariant without verification. Wavelength-resolved characterization and calibration are important when such a beam is used for quantitative multiwavelength polarimetry or birefringence mapping. Further measurements of the polarimeter and the individual optical components are needed to identify the origin of the observed change.

**Funding.** Japan Society for the Promotion of Science (JSPS) (Nos. S20063, JP 23H01416, JP 24H00836, JP 24K03312, JP 25H00395, JP 26H02038)

**Acknowledgment.** The authors are indebted to Dr. Yukihiro Inoue and Mr. Ryo Kurihara (Utsunomiya University) for their helpful technical support and discussions.

**Disclosures.** The authors declare no conflicts of interest.

**Data availability.** Data underlying the results presented in this Letter are not publicly available but may be obtained from the authors upon reasonable request.